\documentclass[11pt]{article}

\newcommand{\beq}{\begin{equation}} 
\newcommand{\eeq}{\end{equation}}
\newcommand{\p}{\partial} 
\newcommand{\bfe}{\mbox{\bf e}}
\newcommand{\no}{\nonumber}

\newcommand{\var}{\varepsilon}

\begin{document}

                          \title{\Large \bf Gravity and the Fermion Mass}

                                   \author{{\large \bf Kenneth Dalton} \\
                                                                       \\
                                    e-mail:  kxdalton@yahoo.com}
\date{}
\maketitle
   
\begin{abstract}
It is shown that gravity generates mass for the fermion.  
It does so by coupling directly with the spinor field.  
The coupling term is invariant with respect to the electroweak gauge group
$ U(1) \otimes SU(2)_L. $ It replaces the fermion mass term $ m\overline{\psi} \psi $.
\end{abstract}

\newpage
\section*{\large {\bf 1. Introduction.}}

The Lagrangian density for the spinor field is derived in appendix A

\beq
  L_f =  \frac{i}{2}\hbar c\biggl\{\overline{\psi} \gamma^\mu \p_\mu \psi 
        - (\p_\mu \overline{\psi})\gamma^\mu \psi \biggr\}
        + \frac{\hbar c}{4} \overline{\psi} \gamma_5 \hat{\gamma}_\delta \psi
          \, \epsilon^{\delta\alpha\beta\gamma} e_\alpha^{\,\,\,\nu} 
             e_\gamma^{\,\,\,\lambda} \p_\lambda e_{\beta\nu}  
\eeq
In this expression, $  \gamma^\mu (x) = e_\alpha^{\,\,\,\mu} (x)\, \hat{\gamma}^\alpha \,$ where 
$\hat{\gamma}^\alpha $ are the constant Dirac matrices.
Gravitation is represented by the tetrad field $ e^\alpha_{\,\,\,\mu}(x) $. 
Because of the coupling between tetrad and spinor, the two fields propagate together.
It is similar to the coupling between electric and magnetic fields 
during the propagation of electromagnetic waves.  

It is shown in appendix B that $ L_f $ yields an electroweak Lagrangian which is invariant 
under the gauge group $ U(1) \otimes SU(2)_L $. The conventional mass term 
$ m\overline{\psi} \psi $ has no such electroweak form.  As a consequence, it cannot appear in the 
electroweak Lagrangian, and it does not contribute to the energy.  
What, then, is the source of fermion mass?  We will find solutions
of the coupling equation which also satisfy the Dirac equation.  Therefore, these solutions
describe particles of mass $m$. 

In this paper, the gauge fields will be ignored altogether.  The gauge interactions
play no part in the generation of fermion mass.

\newpage
\section*{\large {\bf 2. Field Equations.  Energy Tensors.}}
 
The (non-linear) coupling equation follows from $ L_f $ (1) 

\beq
       i \gamma^\mu\p_\mu \psi + 
    \frac{i}{2} \frac{1}{\sqrt{-g}} \p_\mu ( \sqrt{-g} \,\gamma^\mu )\psi  
          +\frac{1}{4} \gamma_5 \hat{\gamma}_\delta \psi 
      \, \epsilon^{\delta\alpha\beta\gamma}e_\alpha^{\,\,\,\nu} 
       e_\gamma^{\,\,\,\lambda} \p_\lambda e_{\beta\nu} = 0
\eeq
This, together with the conjugate equation, yields the conservation law 

\beq
         \p_\mu \left(\sqrt{-g} \, \overline{\psi}\, \gamma^\mu \psi \right) = 0
\eeq
If a plane wave solution exists

\beq
            \psi = \frac{N'}{\sqrt{V}} \left( \begin{array}{c} 
                                                            u_1  \\
                                                            u_2  \\
                                                            u_3 \\
                                                            u_4
                                                       \end{array}
                                                                     \right) \mbox{\rm exp}\,(-i k_\mu x^\mu)
\eeq
then $ \p_\mu \psi = -ik_\mu \psi \;$ and $\; \p_\mu \overline{\psi} = \ ik_\mu \overline{\psi} $.
It follows from the conservation law that  $  \p_\mu (\sqrt{-g}\, \gamma^\mu) = 0, $
and equation (2) simplifies to 

\beq
    i \gamma^\mu\p_\mu \psi   
        +\frac{1}{4} \gamma_5 \hat{\gamma}_\delta \psi 
      \, \epsilon^{\delta\alpha\beta\gamma}e_\alpha^{\,\,\,\nu} 
       e_\gamma^{\,\,\,\lambda} \p_\lambda e_{\beta\nu} = 0
\eeq

The gravitational field equations are 

\beq
    \kappa \, \Big(R_{\mu\nu} - \frac{1}{2} g_{\mu\nu} R \Big) + T_{\mu\nu}^{(m)} = 0
\eeq
where $ \kappa = c^4/8\pi G $ and the Ricci tensor is 

\beq
  R_{\mu\nu} = \p_\nu \Gamma^\lambda_{\mu\lambda} - 
                         \p_\lambda \Gamma^\lambda_{\mu\nu}
                         + \Gamma^\lambda_{\rho\nu} \Gamma^\rho_{\mu\lambda}
                         - \Gamma^\lambda_{\lambda\rho}\Gamma^\rho_{\mu\nu}
\eeq
The material energy tensor is that of the spinor [1]

\begin{eqnarray}
       T_{\mu\nu}^{(m)} &=& \frac{i}{4} \hbar c \Big\{\overline{\psi} \gamma_\mu \p_\nu \psi 
  + \overline{\psi} \gamma_\nu \p_\mu \psi - (\p_\mu \overline{\psi}) \gamma_\nu \psi 
  - (\p_\nu \overline{\psi}) \gamma_\mu \psi \Big\}  \no \\
  & & \hspace{-.6in} + \frac{\hbar c}{4} \overline{\psi} \gamma_5 \hat{\gamma}_\delta \psi 
  \,\epsilon^{\delta\alpha\beta\gamma} e_\gamma^{\,\,\,\lambda} \Big\{ (e_{\alpha\mu} 
  \p_\lambda e_{\beta\nu} + e_{\alpha\nu} \p_\lambda e_{\beta\mu}) 
       - \frac{1}{2} (e_{\alpha\mu}  \p_\nu e_{\beta\lambda} 
                + e_{\alpha\nu} \p_\mu e_{\beta\lambda} )\Big\} \no \\
   &&
\end{eqnarray}
\newpage
The structure of space and time is expressed in terms of a scalar, 3-vector
basis $ e_\mu = (e_0, {\bfe}_i) $.  The basis changes from point to point according to 
the formula 

\beq
        \nabla_{\nu} e_{\mu} = e_{\lambda} Q^{\lambda}_{\mu\nu}
\eeq
which separates into scalar and 3-vector parts

\begin{eqnarray}
      \nabla_\nu e_0 & = & e_0 Q^0_{0 \nu} \\
      \nabla_\nu {\bfe}_i & = & {\bfe}_j Q^j_{i \nu}
\end{eqnarray}
By definition  $ Q^0_{j \nu} = Q^i_{0 \nu} \equiv  0. $  All 28 independent coefficients
$ Q^{\mu}_{\nu\lambda} $ are derivable from the scalar, three-vector metric 
$ g_{\mu\nu} = (g_{00}; g_{ij}). $
Explicitly, 

\begin{eqnarray}
     Q^0_{0\nu} & = & \Gamma^0_{0\nu} = \frac{1}{2} g^{00} \p_\nu g_{00}  \\
     Q^i_{j0} & = & \Gamma^i_{j0} = \frac{1}{2} g^{in} \p_0 g_{nj}  \\
     Q^i_{jk} & = &\Gamma^i_{jk} = \frac{1}{2} g^{in} \left(\p_k g_{jn} + \p_j g_{nk} 
                                                 - \p_n g_{jk}\right)
\end{eqnarray}
where

\beq
   \Gamma^\mu_{\nu\lambda} = \frac{1}{2} g^{\mu\rho}\left(\p_\lambda g_{\nu\rho}
                               + \p_\nu g_{\rho\lambda} - \p_\rho g_{\nu\lambda}\right)
\eeq
are the Christofel coefficients.  The symbols $ \Gamma^\mu_{\nu\lambda} $ are symmetric 
in $ \nu\lambda $, while the $ Q^\mu_{\nu\lambda} $ are not.  The following formula
holds good 

\beq
      Q^\mu_{\nu\lambda} = \Gamma^\mu_{\nu\lambda} +
                         g^{\mu\rho}  g_{\lambda\eta}  Q^\eta_{[\nu\rho]}
\eeq
where 

\beq
        Q^\mu_{[\nu\lambda]} \equiv  Q^\mu_{\nu\lambda} - Q^\mu_{\lambda\nu}
\eeq

\newpage
The gravitational energy tensor is given by

\beq
  T^{(g)}_{\mu\nu} = \kappa \,\Big\{Q^\rho_{[\lambda\mu]} Q^\lambda_{[\rho\nu]} + Q_\mu Q_\nu 
    - \frac{1}{2} g_{\mu\nu} g^{\eta\tau}(Q^\rho_{[\lambda\eta]}Q^\lambda_{[\rho\tau]}  
    + Q_\eta Q_\tau)\Big\}
\eeq
where $ Q_\mu = Q^\rho_{[\rho\mu]}. $ There are 9 independent components $ Q^\mu_{[\nu\lambda]},$ namely

\beq
   Q^0_{0i} = \frac{1}{2} g^{00} \p_i g_{00} \;\;\;\; \mbox{\rm and}\;\;\;\;
   Q^i_{j0} = \frac{1}{2} g^{in} \p_0 g_{nj}
\eeq
For a static, Newtonian potential $\psi$  

\beq
       g_{00} = 1 + \frac{2}{c^2} \psi 
\eeq
so that
 
\beq
         Q^0_{0i} = \frac{1}{c^2} \p_i \psi \;\;\;\; \mbox{\rm and}\;\;\;\;
                      Q^i_{j0} = 0
\eeq
It follows that

\begin{eqnarray}
    T^{(g)}_{00} & = & \frac{1}{8\pi G} (\nabla \psi)^2 \\
    T^{(g)}_{0i} & = & 0 \\
    T^{(g)}_{ij} & = & \frac{1}{4\pi G} \Big\{\p_i \psi \, \p_j \psi
         - \frac{1}{2} \delta_{ij} (\nabla \psi)^2 \Big\}
\end{eqnarray}
which is the Newtonian stress-energy tensor.   
The topic of energy, momentum, and stress is taken up in section 6. 

In the calculations to follow, frequent use is made of the conditions 
\footnote{The components $ e^a_{\,\,\,i} $ are projections of
the basis vectors $ \bfe_i $ onto an orthonormal frame.  The $ e^a_{\,\,\,i} $ can be made
symmetric, at any point, by suitably rotating the frame.  The three rotation parameters yield
the three conditions (25).  These frame rotations do not affect the coordinate system $ \{x^\mu\} $.}

\beq
    e^1_{\,\,\,2} = e^2_{\,\,\,1} \;\;\;\; e^2_{\,\,\,3} = e^3_{\,\,\,2} \;\;\;\;
        e^3_{\,\,\,1} = e^1_{\,\,\,3}
\eeq
Moreover, the focus is upon oscillating gravitational fields of very small amplitude. 
The coordinate system is taken to be nearly rectangular 

\begin{eqnarray}
             e^\alpha_{\,\,\,\mu} & = & \delta^\alpha_{\,\,\,\mu} + \xi^\alpha_{\,\,\,\mu}  \\
             e_\alpha^{\,\,\,\mu} & = & \delta_\alpha^{\,\,\,\mu} - \xi_\alpha^{\,\,\,\mu}  \\
      e^\alpha_{\,\,\,\mu}\, e_\alpha^{\,\,\,\nu} & = & \delta^\nu_\mu \,
             =  \,(\delta^\alpha_{\,\,\, \mu} + \xi^\alpha_{\,\,\, \mu})
                             (\delta_\alpha^{\,\,\, \nu} - \xi_\alpha^{\,\,\, \nu})  \no \\
                   & = & \delta^\nu_\mu + \delta_\alpha^{\,\,\, \nu}\xi^\alpha_{\,\,\, \mu}
                                     - \delta^\alpha_{\,\,\, \mu}\xi_\alpha^{\,\,\, \nu}  + O(\xi^2) 
\end{eqnarray}
Therefore, 

\beq
   \xi^\nu_{\;\; \mu}  =  \xi_\mu^{\;\; \nu} + O(\xi^2)
\eeq
($ \xi^\alpha_{\;\;\mu} $ and $ \xi_\alpha^{\;\;\mu} $ are small compared with unity.)
Only the largest terms will be retained in the Ricci tensor 

\beq
  R_{\mu\nu} = \p_\nu \Gamma^\lambda_{\mu\lambda} - 
                         \p_\lambda \Gamma^\lambda_{\mu\nu}
\eeq
Expansion of the metric

\beq
  g_{\mu\nu} = \eta_{\alpha\beta} \, e^\alpha_{\;\; \mu} e^\beta_{\;\; \nu} = 
   \eta_{\alpha\beta} (\delta^\alpha_{\;\; \mu} + \xi^\alpha_{\;\; \mu})
             (\delta^\beta_{\;\; \nu} + \xi^\beta_{\;\; \nu})
\eeq
yields

\beq
   R_{\mu\nu} =  \eta^{\lambda\rho} \p_\lambda \p_\rho \xi_{\mu\nu} 
    + \p_\mu \p_\nu \xi^\lambda_{\;\; \lambda}  
       - \p_\mu \p_\lambda \xi^\lambda_{\;\; \nu} - \p_\nu \p_\lambda \xi^\lambda_{\;\; \mu}
\eeq
and  

\beq
       R = g^{\mu\nu} R_{\mu\nu} = 2 \eta^{\mu\nu} (\p_\mu \p_\nu \xi^\lambda_{\;\; \lambda}
             - \p_\mu \p_\lambda \xi^\lambda_{\;\; \nu})
\eeq

\newpage
\section*{\large {\bf 3. The Coupling Equation.}}
 
Define the pseudovector

\beq
     \var^\delta = - \frac{1}{4} \epsilon^{\delta\alpha\beta\gamma}e_\alpha^{\,\,\,\nu} 
       e_\gamma^{\,\,\,\lambda} \p_\lambda e_{\beta\nu}
\eeq
so that the coupling equation (5) becomes 

\beq
    i \gamma^\mu\p_\mu \psi -  \gamma_5 \hat{\gamma}_\delta  \var^\delta \psi = 0
\eeq
At the end of this section, it will be shown that tetrad solutions $ e^\alpha_{\;\;\mu}(x) $ exist
such that $ \var^\delta $ is constant in space and time.  Anticipating this, 
substitute the trial solution (4) 

\beq
     (k_\alpha \hat{\gamma}^\alpha   - \var_\alpha \gamma_5 \hat{\gamma}^\alpha) \psi = 0
\eeq
where $  k_\alpha = e_\alpha^{\;\;\mu}  k_\mu $.  Make use of the matrix representation 

\beq
         \hat{\gamma}^0 = \left(\!\! \begin{array}{cc}
                                                     \sigma_0 &0 \\
                                                      0& -\sigma_0
                                               \end{array}
                                               \! \!\right) 
   \;\;\;\; \hat{\gamma}^a = \left(\!\! \begin{array}{cc}
                                                     0& \sigma_a \\
                                                    -\sigma_a &0 
                                                \end{array}
                                                  \!\! \right)  
   \;\;\;\;  \gamma_5 = \left(\!\! \begin{array}{cc}
                                           0 & 1  \\
                                           1 & 0 
                                          \end{array}
                                            \! \! \right)
\eeq
and then introduce 
     $ u =   N   \left(\!\! \begin{array}{c}
                \phi  \\
                \chi
             \end{array} \!\!\right) $
to find

\beq
                            \left(\!\! \begin{array}{cc}                        
                            (k^0 - \sigma_a \var^a)   & (\var^0 -\sigma_a k^a) \\         
                             -(\var^0 -\sigma_a k^a) & - (k^0 -\sigma_a \var^a)            
                             \end{array} \!\!\right)
                               \left(\!\! \begin{array}{c}
                                   \phi  \\
                                    \chi
                               \end{array} \!\!\right) = 0
\eeq
or

\begin{eqnarray}
  (k^0 - \sigma_a \var^a) \phi +  (\var^0 -\sigma_a k^a) \chi & = & 0  \\
  (\var^0 -\sigma_a k^a) \phi +  (k^0 - \sigma_a \var^a) \chi & = & 0
\end{eqnarray}
Add and subtract, in order to arrive at the uncoupled equations 

\begin{eqnarray}
   \{(k^0 + \var^0) - \sigma_a (k^a + \var^a)\}(\phi + \chi) & = & 0  \\
   \{(k^0 - \var^0) + \sigma_a (k^a - \var^a)\}(\phi - \chi) & = & 0 
\end{eqnarray} 
For each equation, the determinant of the coefficient must vanish.  In the first case,

\beq
   \left| \begin{array}{cc} 
       (k^0 + \var^0) - (k^3 + \var^3) & -(k_- + \var_-)  \\
      -(k_+ + \var_+)   &   (k^0 + \var^0) + (k^3 + \var^3) 
    \end{array} \right| = 0 
\eeq
where $ k_\pm = k^1 \pm ik^2 \,$ and $\, \var_\pm = \var^1 \pm i\var^2 $.
This gives

\beq 
    k_\mu k^\mu + \var_\mu \var^\mu + 2k_\mu \var^\mu = 0
\eeq
while the second equation gives

\beq 
    k_\mu k^\mu + \var_\mu \var^\mu - 2k_\mu \var^\mu = 0
\eeq
Therefore, the constants $ k^\mu $ and $ \var^\mu $ are related by

\beq
     k_\mu \var^\mu = 0  \hspace{.25in} \mbox{\rm and} 
         \hspace{.25in} k_\mu k^\mu + \var_\mu\var^\mu = 0
\eeq

\begin{center}
               -------------------------------
\end{center}

In the expansion of $ \var^\delta $, all first order terms vanish

\begin{eqnarray}
 \var^\delta & = & - \frac{1}{4} \epsilon^{\delta\alpha\beta\gamma}e_\alpha^{\,\,\,\nu} 
       e_\gamma^{\,\,\,\lambda} \p_\lambda e_{\beta\nu} 
             = - \frac{1}{4} \epsilon^{\delta ab\gamma}e_a^{\,\,\,n} 
       e_\gamma^{\,\,\,\lambda} \p_\lambda e_{bn}    \no \\
             & = &  \frac{1}{4} \epsilon^{\delta ab\gamma} \xi_a^{\,\,\,n} 
                      \p_\gamma \xi_{bn}    
\end{eqnarray}
This expression contains no derivatives of $ e^0_{\;\;0} $.  Thus, it plays no
role in the spinor coupling, $ e^0_{\;\;0} = 1 $.  With the substitution of the 
trial solution  

\beq
    \xi^i_{\;\;j} = a^i_{\;\;j} \cos(-K_\mu x^\mu) + b^i_{\;\;j} \sin(-K_\mu x^\mu)
\eeq
$  \var^\delta $ becomes constant in space and time (as was to be shown)

\beq
    \var^\delta  =  -\frac{1}{4} \epsilon^{\delta ab\gamma} K_\gamma \,a_{an} b^n_{\;\;b} 
\eeq
For later use, the components are

\begin{eqnarray}
   \var^c & = & -\frac{1}{4} \epsilon^{cab0} K_0 \, a_{an} b^n_{\;\;b} 
            =  \frac{1}{4} \epsilon^{0abc} K_0 \,a_{an} b^n_{\;\;b}   \\
   \var^0 & = & -\frac{1}{4} \epsilon^{0abc} K_c \,a_{an} b^n_{\;\;b}  
               = -\frac{K_c}{K_0} \,\var^c   
\end{eqnarray}
or $ K_\mu \var^\mu = 0. $

\newpage

\section*{\large {\bf 4. The Dirac Equation.  Helicity.}}

Fermions are described by solutions of the Dirac equation [2]

\beq
  i \gamma^\mu\p_\mu \psi - m \psi = 0
\eeq 
[In this section, $ \hbar = c = 1 $.  The replacement $m \longrightarrow mc/\hbar $
gives the explicit formula.]  
Substitute the plane wave (4) with $ u = N \left(\!\! \begin{array}{c}
                             \phi  \\
                               \chi
                   \end{array} \!\!\right)  $ to find

\beq
   \chi = \frac{\sigma_a k^a}{k^0 + m}\phi \hspace{.25in} \mbox{\rm and} \hspace{.25in} k_\mu k^\mu = m^2
\eeq
The two component spinor is normalized to unity, $\phi^\dagger \phi = 1, $
while $u$ is normalized to $ k^0/m $ 

\beq
    u^\dagger u = N^2 \frac{2k^0}{k^0 + m} = \frac{k^0}{m}  \hspace{.5in} N = \sqrt{\frac{k^0 + m}{2m}}
\eeq

In the following section, it will be shown that functions (48)
satisfy the gravitational field equations, if the spinor is in a state of definite 
helicity.  Helicity states are defined by the eigenvalue equation [3, 4, 5]

\beq
     \frac{\sigma_a k^a}{k} \phi_\pm = \pm \phi_\pm 
\eeq
where $ k = [(k^1)^2 + (k^2)^2 + (k^3)^2]^{1/2} .$
For positive helicity,    

\beq
  \left(\!\! \begin{array}{cc}                        
              (k^3 - k)  & k_- \\         
              k_+  & -(k^3 + k)             
                  \end{array} \!\!\right) 
       \left(\!\! \begin{array}{c}
                \phi_1  \\
                \phi_2
             \end{array} \!\!\right)    = 0                
\eeq
which yields $ \phi_2 = \frac{k_+}{k + k^3}\, \phi_1 $.  Normalize $\phi^\dagger \phi = 1 $
to find $ \phi_1 = \sqrt{\frac{k + k^3}{2k}} $ and 

\beq
    \phi_+ = \frac{1}{\sqrt{2k(k + k^3)}}\left(\!\! \begin{array}{c}
                k + k^3  \\
                k_+ 
             \end{array} \!\!\right)  
\eeq
Similarly,

\beq
    \phi_- = \frac{1}{\sqrt{2k(k + k^3)}}\left(\!\! \begin{array}{c}
                -k_-   \\
                k + k^3 
             \end{array} \!\!\right)  
\eeq
If the motion is along  $ x^3$, then these eigenstates coincide with those of spin, 
$\, \phi_+ = \left(\!\! \begin{array}{c}
                1  \\
                0 
             \end{array} \!\!\right) $ and 
$ \phi_- = \left(\!\! \begin{array}{c}
                0  \\
                1 
             \end{array} \!\!\right) $.

For the remainder of the paper, we focus upon the positive frequency, positive helicity solutions
of the Dirac equation

\beq
  \psi_+  =  \sqrt{\frac{m}{k^0 V}} \, u_+ \exp (-ik_\mu x^\mu)
\eeq
where

\beq
        u_+ = N \left(\!\!\! \begin{array}{c} \phi_+ \\  
             \frac{k}{k^0 + m}\phi_+ \end{array} \!\!\!\right) 
\eeq
The solution $ \psi_+ $ must also satisfy the coupling equation (35).  This will be the case, if 

\beq
      \gamma_5 \hat{\gamma}_\delta \,\var^\delta \, u_+  = m u_+
\eeq
The corresponding two-component systems are 

\begin{eqnarray}
  \biggl\{\sigma_a\var^a - m - \frac{\var^0 k}{k^0 + m} \biggr\} \left(\!\! \begin{array}{c}
                k + k^3  \\
                k_+ 
             \end{array} \!\!\right) & = & 0   \\
         \biggl\{\var^0 - \frac{k(\sigma_a\var^a + m)}{k^0 + m}\biggr\}  
          \left(\!\! \begin{array}{c}
                     k + k^3  \\
                     k_+ 
                     \end{array} \!\!\right) & = & 0 
\end{eqnarray}
After some algebra, these equations yield the expression           

\beq
    \var^{a} = \frac{k^a}{k}\left(\frac{\var^0 k}{k^0 + m}  + m\right)  
\eeq
Make use of $ k_\mu \var^\mu = 0 $ \,(46) in order to obtain            
 
\beq
  \var^0 = k \hspace{.2in} \mbox{\rm and } \hspace{.2in} \var^a = \frac{k^a k^0}{k} \hspace{.3in} a = 1,2,3
\eeq
This solution, together with (49), relates the propagation
vectors $ k_\mu $ and $ K_\mu $.  In appendix C, it is shown that $ k_\mu = \alpha K_\mu $,
where $\alpha$  is a very small constant, second order in the amplitudes.  Therefore,
the spinor and tetrad propagate with the same phase velocity $ k^0/k^i = K^0/K^i $.

\newpage
\section*{\large {\bf 5.  Gravity.}}

It remains to be determined whether the gravitational field equations are satisfied. 
Setting $ \xi^0_{\;\; 0} = 0 $ in (32) leaves 

\begin{eqnarray}
     R_{00} & = & \p_0 \p_0 \xi^n_{\;\; n}  \\
     R_{ij} & = &  \eta^{\lambda\rho} \p_\lambda \p_\rho \xi_{ij} + \p_i \p_j \xi^n_{\;\; n}  
       - \p_i \p_n \xi^n_{\;\; j} - \p_j \p_n \xi^n_{\;\; i}
\end{eqnarray}
Conditions $ \p_\mu (\sqrt{-g}\,e_\alpha^{\;\;\mu} )\,\hat{\gamma}^\alpha = 0 $
(section 2) are $ \p_\lambda \xi_\alpha^{\;\;\lambda} 
+ \p_\alpha \xi^\lambda_{\;\;\lambda} = 0 $ to first order.  It follows that 

\beq
   \xi^n_{\;\; n} = 0 \hspace{.2in} \mbox{\rm and } \hspace{.2in} \p_n \xi^n_{\;\; i} = 0
               \hspace{.3in} i = 1,2,3 
\eeq
Therefore, the solution is both traceless and transverse; also $ R_{00} = R = 0 $ and 

\beq
    R_{ij} =  \eta^{\lambda\rho} \p_\lambda \p_\rho \xi_{ij} 
\eeq
Since $ R_{ij} $ is first order in $ \xi_{ij} $, only first order terms will be retained
in $ T_{\mu\nu}^{(m)} \;$(8). The derivative $ \p_\mu \psi $ is proportional to $ k_\mu $, 
which was shown to be second order in $\, \xi_{ij} $.  Expand $ T_{\mu\nu}^{(m)} $ to find  
$ T_{00}^{(m)} = 0 $ and

\beq
    T_{ij}^{(m)} =  \frac{\hbar c}{4} \overline{\psi} \gamma_5 \hat{\gamma}_\delta \psi 
  \,\epsilon^{\delta ab \gamma} (\eta_{ai} \p_\gamma \xi_{bj} + \eta_{aj} \p_\gamma \xi_{bi}) 
\eeq
The factor  $ \overline{\psi} \gamma_5 \hat{\gamma}_\delta \psi $ is evaluated
by means of the helicity state 

\beq
   \overline{\psi}_+ \gamma_5 \hat{\gamma}_\delta \psi_+ = 
         \frac{mc}{\hbar k^0 V} \,\overline{u}_+ \gamma_5 \hat{\gamma}_\delta u_+ 
             = - \frac{\var_\delta}{k^0 V} 
\eeq
therefore

\beq
    T_{ij}^{(m)} =  -\frac{\hbar c}{4 k^0 V} \,\epsilon^{\delta ab \gamma} \var_\delta 
               (\eta_{ai} \p_\gamma \xi_{bj} + \eta_{aj} \p_\gamma \xi_{bi}) 
\eeq
Substitute the trial solution (48) and simplify, making use of  $ k_\mu= \alpha K_\mu $, 
in order to obtain 

\begin{eqnarray}
    T_{ij}^{(m)} &\!\!\! =\!\!\! & \frac{\hbar c}{4 k^0 V} (K_0^2 - K^2) \,\epsilon^{0abc} 
             \frac{K_c}{K} \Big\{\eta_{ai} a_{bj} \sin (-K_\mu x^\mu)
                    - \eta_{ai} b_{bj} \cos (-K_\mu x^\mu)\Big\}   \no \\
           && + \;\; i \leftrightarrow j
\end{eqnarray}

\newpage
At this point, the gravitational field equations are 

\beq
     -\kappa R_{ij} =  T_{ij}^{(m)} 
\eeq
The integration volume is introduced on the left-hand side by means of a 
length parameter $ \lambda $: $ \kappa \longrightarrow \kappa \lambda^3/V $.  
Substitute the trial solution to find 

\begin{eqnarray}
    -\kappa R_{ij} & = & \frac{\kappa \lambda^3}{V} \,(K_0^2 - K^2)\,\xi_{ij} \no \\                   
                   & = & \frac{\kappa \lambda^3}{V}\,(K_0^2 - K^2) 
               \Big\{a_{ij} \cos (-K_\mu x^\mu) + b_{ij} \sin (-K_\mu x^\mu)\Big\} 
\end{eqnarray}  
Equate coefficients of $ \cos (-K_\mu x^\mu) $ and $ \sin (-K_\mu x^\mu) $
in (73) and (75) to find that solutions exist only if 

\beq
    \kappa \lambda^3 = \frac{\hbar c}{2 K^0}
\eeq
In this case, 

\begin{eqnarray}
      a_{ij} & = & - \frac{1}{2}\,\epsilon^{0abc} \frac{K_c}{K} 
                   \Big\{\eta_{ai} b_{bj} + \eta_{aj} b_{bi}\Big\}  \\
      b_{ij} & = &  \frac{1}{2}\,\epsilon^{0abc} \frac{K_c}{K} 
                   \Big\{\eta_{ai} a_{bj} + \eta_{aj} a_{bi}\Big\}
\end{eqnarray}
Therefore, the gravitational field equations are satisfied.  An explicit solution
is given in appendix D.  The following are two special cases:  


(a)  Motion along the $x^3$-axis; $k^1 = k^2 = K^1 = K^2 = 0$. \newline
\noindent
In this case, $ K = K^3 = -K_3. $ From appendix D, $ a^3_{\;\; i} = b^3_{\;\; i} = 0 $.
 The solution is 

\begin{eqnarray}
    \xi^1_{\;\; 1} & = & a^1_{\;\; 1} \cos(-K_\mu x^\mu) + b^1_{\;\; 1} \sin(-K_\mu x^\mu) \\
    \xi^2_{\;\; 2} & = & - \xi^1_{\;\; 1}   \\
    \xi^1_{\;\; 2} & = & -b^1_{\;\; 1} \cos(-K_\mu x^\mu) + a^1_{\;\; 1} \sin(-K_\mu x^\mu) 
\end{eqnarray}
If the phase is such that $ b^1_{\;\; 1} = 0, $ then this is identical to the solution found in [1].

\newpage
(b)  Motion in the $(x^2 ,x^3)$  plane;  $k^1 = K^1 = 0$. \newline
\noindent
In this case, $ (K)^2 = K_2^2 + K_3^2 $.  The solution is
  
\begin{eqnarray}
    \xi^1_{\;\; 1} & = & a^1_{\;\; 1} \cos(-K_\mu x^\mu) + b^1_{\;\; 1} \sin(-K_\mu x^\mu) \\
    \xi^2_{\;\; 2} & = & - \left(\frac{K_3}{K}\right)^2 \biggl\{a^1_{\;\; 1} \cos(-K_\mu x^\mu) +
                                b^1_{\;\; 1} \sin(-K_\mu x^\mu) \biggr\} \\
    \xi^3_{\;\; 3} & = & - \left(\frac{K_2}{K}\right)^2 \biggl\{a^1_{\;\; 1} \cos(-K_\mu x^\mu) +
                                b^1_{\;\; 1} \sin(-K_\mu x^\mu)\biggr\} \\
    \xi^1_{\;\; 2} & = &  \frac{K_3}{K} \biggl\{b^1_{\;\; 1} \cos(-K_\mu x^\mu) -
                                a^1_{\;\; 1} \sin(-K_\mu x^\mu)\biggr\} \\
    \xi^2_{\;\; 3} & = &  \frac{K_2 K_3}{(K)^2} \biggl\{a^1_{\;\; 1} \cos(-K_\mu x^\mu) +
                                b^1_{\;\; 1} \sin(-K_\mu x^\mu)\biggr\} \\
    \xi^3_{\;\; 1} & = & - \frac{K_2}{K} \biggl\{b^1_{\;\; 1} \cos(-K_\mu x^\mu) -
                                a^1_{\;\; 1} \sin(-K_\mu x^\mu)\biggr\}
\end{eqnarray}

\newpage
\section*{\large {\bf 6.  Energy, Momentum, Stress. }}

The total density of energy, momentum, and stress is given by 

\beq
    T_{\mu\nu} = T_{\mu\nu}^{(g)} + T_{\mu\nu}^{(m)}
\eeq 
The gravitational energy tensor (18) is simplified in the present case, because 
$ Q^0_{0i} = \p_i \xi^0_{\;\;0} = 0 \;$ and $\;  Q^n_{n0} = \p_0 \xi^n_{\;\;n} = 0 $,
so that $ Q_\mu = 0 $.  This leaves 

\begin{eqnarray}
    T^{(g)}_{00} & = & \frac{\hbar c}{4 K^0 V} \, Q^l_{m0}Q^m_{l0}     \\
    T^{(g)}_{0i} & = & 0                                              \\
    T^{(g)}_{ij} & = & \frac{\hbar c}{4 K^0 V} \, Q^l_{m0}Q^m_{l0}\, \delta_{ij}         
\end{eqnarray}
Make use of (77, 78) to find

\beq
   a^{lm} a_{lm} = b^{lm} b_{lm} = -\epsilon^{0abc} \frac{K_c}{K} a_{an} b^n_{\;\;b}
\eeq
while $ a^{lm} b_{lm} = 0 $.  It follows that 

\beq
   Q^l_{m0}Q^m_{l0} = \p_0 \xi^{lm} \p_0 \xi_{lm} =  (K_0)^2 \,a^{lm} a_{lm} 
\eeq
Compare (92) with (51), then make use of $ \var^0 = k \,$ and $\, k_\mu = \alpha K_\mu $ 
in order to obtain $ a^{lm} a_{lm}  = 4 \alpha .$  Therefore, $ Q^l_{m0}Q^m_{l0} = 4 k_0 K_0 \,$ and

\begin{eqnarray}
    T^{(g)}_{00} & = & \frac{\hbar c k_0}{V} = \frac{\hbar \omega}{V}  \\
    T^{(g)}_{0i} & = & 0 \\
    T^{(g)}_{ij} & = & \frac{\hbar \omega}{V} \,\delta_{ij}         
\end{eqnarray}
The stresses are constant and purely compressive.  

With regard to the matter tensor $T_{\mu\nu}^{(m)}$, the stress components $T_{ij}^{(m)}$ were given in (73).
They are first order and propagate along with the tetrad.  Components  $T_{00}^{(m)}$ and $T_{0i}^{(m)}$ 
are calculated from (8) by means of 

\beq 
\overline{\psi}_+ \hat{\gamma}_\delta \, \psi_+ =  \frac{k_\delta}{k^0 V} \;\;\;\; \mbox{\rm and} \;\;\;\; 
\overline{\psi}_+ \gamma_5 \hat{\gamma}_\delta \,\psi_+ = -\frac{\var_\delta}{k^0 V}
\eeq
The material energy density is 

\beq
   T_{00}^{(m)}  =  \frac{\hbar c k_0}{V} - \frac{\hbar c}{4 k_0 V} 
                  \var_a \epsilon^{0abc} \xi_b^{\;\;n} \p_0 \xi_{cn}  
                 = \frac{\hbar c k_0}{V} + \frac{\hbar c}{k_0 V} \var_a \var^a \;=\; 0 
\eeq
where (49) and (65) have been used. 
The density of momentum is 

\beq
   T_{0i}^{(m)}  =  \frac{\hbar c k_i}{V} - \frac{\hbar c}{4 k_0 V} \var_a \epsilon^{0abc} 
                    e_b^{\;\;n} \p_n e_{ci}  
              + \frac{\hbar c}{8 k_0 V} \epsilon^{0abc}  
             \xi_a^{\;\;n} \Big\{\var_i \, \p_c \xi_{bn} - \var_c \,\p_i \xi_{bn} \Big\} 
\eeq
Substitute the solution (48) to find that the last term is zero.  The 
second term is a divergence $\, e_b^{\;\;n} \p_n e_{ci}  = \p_n (e_b^{\;\;n} e_{ci}) ,\,$ since
$ \, \p_n e_b^{\;\;n} = 0 $.  Therefore, 

\beq
   T_{0i}^{(m)}  =  \frac{\hbar c k_i}{V} \; + \;\;\mbox{\rm a divergence}
\eeq
In sum, the fermion's energy is given by the gravitational component 
$ T^{(g)}_{00} $, while its momentum is given by $ T_{0i}^{(m)}  $.
The total energy $\hbar \omega$ is conserved, as is the total momentum $\hbar k^i$.

\newpage
\section*{\large {\bf 7. Concluding Remarks.}}

Despite the equality (61), the coupling equation and the Dirac equation 
are not equivalent.  They differ in their matrix algebra.  
The coupling equation conserves not only the vector current (3), 
but the axial vector current as well

\beq
         \p_\mu \left(\sqrt{-g} \, \overline{\psi}\, \gamma_5 \gamma^\mu \psi \right) = 0
\eeq
The Dirac equation does not conserve the axial current 

\beq
         \p_\mu \left(\sqrt{-g} \, \overline{\psi}\, \gamma_5 \gamma^\mu \psi \right) = 
                 - 2im \sqrt{-g} \, \overline{\psi}\, \gamma_5  \psi 
\eeq
This is clearly due to the mass term $ m \overline{\psi} \psi $.
The electroweak interaction contains conserved vector and axial vector parts.  Thus, 
the mass term cannot appear in the Lagrangian.

\newpage
\section*{\large {\bf Appendix A: Spinor Lagrangian.}}

The covariant spinor derivative is [6, 7] 

\beq
      D_\mu \psi = \p_\mu \psi + \Gamma_\mu \psi
\eeq
where

\beq
 \Gamma_\mu = \frac{1}{8} \left(\hat{\gamma}^\alpha \hat{\gamma}^\beta 
    - \hat{\gamma}^\beta \hat{\gamma}^\alpha \right) \omega_{\alpha\beta\mu} 
  = \frac{1}{4} \hat{\gamma}^{[\alpha} \hat{\gamma}^{\beta ]} \,\omega_{\alpha\beta\mu}
\eeq
The conjugate expression is  

\beq
     D_\mu\overline{\psi} = \p_\mu \overline{\psi} - \overline{\psi} \,\Gamma_\mu 
\eeq
The spinor Lagrangian density is given by

\begin{eqnarray} 
    L_f & = & \frac{i}{2} \hbar c \Big\{ \overline{\psi} \gamma^\mu D_\mu \psi 
             - (D_\mu \overline{\psi}) \gamma^\mu \psi \Big\}    \no  \\
      & = &  \frac{i}{2} \hbar c \Big\{ \overline{\psi} \gamma^\mu \p_\mu \psi 
             - (\p_\mu \overline{\psi}) \gamma^\mu \psi \Big\} 
      + \, \frac{i}{2} \hbar c \overline{\psi}\left(\gamma^\mu \Gamma_\mu 
               + \Gamma_\mu \gamma^\mu \right) \psi 
\end{eqnarray}
The coupling term may be reduced, by first expanding

\begin{eqnarray}
   \gamma^\mu \Gamma_\mu  + \Gamma_\mu \gamma^\mu & = & 
       \frac{1}{4} e_\gamma^{\,\,\,\mu} \left(\hat{\gamma}^\gamma \hat{\gamma}^{[\alpha}
      \hat{\gamma}^{\beta ]} +  \hat{\gamma}^{[\alpha} \hat{\gamma}^{\beta ]} 
          \hat{\gamma}^\gamma \right) \omega_{\alpha\beta\mu}  \no  \\
  & = & \frac{1}{2} \hat{\gamma}^{[\alpha} \hat{\gamma}^\beta 
           \hat{\gamma}^{\gamma ]} \omega_{\alpha\beta\gamma}
\end{eqnarray}
where

\beq
        \hat{\gamma}^{[\gamma} \hat{\gamma}^\alpha \hat{\gamma}^{\beta ]}  
            =  \frac{1}{2}  \left( \hat{\gamma}^\gamma \hat{\gamma}^{[\alpha}
            \hat{\gamma}^{\beta ]} +  \hat{\gamma}^{[\alpha} \hat{\gamma}^{\beta ]} 
                    \hat{\gamma}^\gamma \right)
\eeq
The identity [7]

\beq
        \hat{\gamma}^{[\alpha} \hat{\gamma}^\beta \hat{\gamma}^{\gamma ]}  
     \equiv    -i \gamma_5 \hat{\gamma}_\delta \,\epsilon^{\delta\alpha\beta\gamma}
\eeq
then gives

\begin{eqnarray} 
    \gamma^\mu \Gamma_\mu  + \Gamma_\mu \gamma^\mu & = & 
          -\frac{i}{2} \gamma_5 \hat{\gamma}_\delta \,\epsilon^{\delta\alpha\beta\gamma}
            \omega_{\alpha\beta\gamma}                       \no  \\
          & = & -\frac{i}{2} \gamma_5 \hat{\gamma}_\delta \,\epsilon^{\delta\alpha\beta\gamma}
            \omega_{[\alpha\beta\gamma]}
\end{eqnarray}
since $ \epsilon^{\delta\alpha\beta\gamma} $ is totally anti-symmetric. 

The covariant derivative of the tetrad is zero 

\beq
    \p_\nu e_\alpha^{\;\;\mu} - e_\beta^{\;\;\mu} \omega^\beta_{\;\;\alpha\nu} 
            + e_\alpha^{\;\;\lambda} Q^\mu_{\;\;\lambda\nu} = 0
\eeq 
so that 

\beq
      Q^\mu_{\;\;\lambda\nu}  =  e_\alpha^{\;\;\mu}  \p_\nu e^\alpha_{\;\;\lambda} 
                              + \omega^\mu_{\;\;\lambda\nu}  
\eeq
Beginning with formula (16), it is readily shown that the anti-symmetric tensor

\beq
         Q_{[ \mu\nu\lambda ]} = g_{\mu\rho} Q^\rho_{[ \nu\lambda ]} 
        + g_{\nu\rho} Q^\rho_{[ \lambda\mu ]} + g_{\lambda\rho} Q^\rho_{[ \mu\nu ]} = 0
\eeq
Therefore,

\beq
         0 = e^\alpha_{\,\,[ \rho} \p_\nu e_{\alpha\mu ]} + \omega_{ [ \rho\mu\nu ]}
\eeq
or

\beq
       \omega_{ [\mu\nu\lambda] } = e^\alpha_{\,\,[ \mu} \p_\nu e_{\alpha\lambda ]}
\eeq
Substitute this expression into (110) and (106), in order to obtain

\beq
  L_f =  \frac{i}{2}\hbar c\biggl\{\overline{\psi} \gamma^\mu \p_\mu \psi 
        - (\p_\mu \overline{\psi})\gamma^\mu \psi \biggr\}
        + \frac{\hbar c}{4} \overline{\psi} \gamma_5 \hat{\gamma}_\delta \psi
          \, \epsilon^{\delta\alpha\beta\gamma} e_\alpha^{\,\,\,\nu} 
             e_\gamma^{\,\,\,\lambda} \p_\lambda e_{\beta\nu}  
\eeq

\clearpage
\section*{\large {\bf Appendix B. \boldmath{$ U(1) \otimes SU(2)_L$ } Gauge Invariance. }}

The gravitational coupling term in $ L_f $ (1) contains the factor 
$ \overline{\psi} \gamma_5 \hat{\gamma}_\delta \psi $.  This factor does not mix
right- and left-handed spinor components, $ \psi_R $ and $ \psi_L $.  In order to
prove this, set

\beq
 \psi = \psi_R + \psi_L = \frac{1 + \gamma_5}{2}\, \psi + 
     \frac{1 - \gamma_5}{2} \,\psi
\eeq
where $ (\gamma_5)^2 = 1 $ and $ \gamma_5 \gamma_\delta = - \gamma_\delta \gamma_5 $.
Also, $ \overline{\psi}_R = \overline{\psi}\,\,\frac{1 - \gamma_5}{2}\, $ and 
$\, \overline{\psi}_L = \overline{\psi}\,\,\frac{1 + \gamma_5}{2} $.\newline
In the expansion

\begin{eqnarray}
\overline{\psi} \gamma_5 \hat{\gamma}_\delta \psi & = &
  (\overline{\psi}_R + \overline{\psi}_L )\, \gamma_5 \gamma_\delta \,
(\psi_R + \psi_L) \no \\
  & = & \overline{\psi}_R \gamma_5 \gamma_\delta \psi_R 
         + \overline{\psi}_L \gamma_5 \gamma_\delta \psi_L 
         + \overline{\psi}_R \gamma_5 \gamma_\delta \psi_L 
         + \overline{\psi}_L \gamma_5 \gamma_\delta \psi_R 
\end{eqnarray}
the mixed terms are identically zero.  For example,

\beq
  \overline{\psi}_R \gamma_5 \gamma_\delta \psi_L =  
   \overline{\psi}\,\frac{1-\gamma_5}{2} \,\gamma_5 \gamma_\delta 
           \,\frac{1-\gamma_5}{2}\,\psi
  = \overline{\psi}\,\gamma_5 \gamma_\delta 
           \,\frac{1-(\gamma_5)^2}{4} \,\psi = 0
\eeq
Therefore, 

\beq
  \overline{\psi} \gamma_5 \hat{\gamma}_\delta \psi = 
    \overline{\psi}_R \gamma_5 \gamma_\delta \psi_R 
         + \overline{\psi}_L \gamma_5 \gamma_\delta \psi_L
\eeq

An expression of this type will be invariant under 
$ U(1) \otimes SU(2)_L $ gauge transformations.
\footnote{The Dirac mass term $\, m \overline{\psi} \psi =
m(\overline{\psi}_R {\psi}_L + \overline{\psi}_L {\psi}_R)\, $
mixes right- and left-handed spinors and cannot appear in
the electroweak Lagrangian.}
Introduce the right-handed singlet $ \psi_R = e_R $ and
left-handed doublet $\psi_L = \left(\begin{array}{c} \nu_L \\
e_L \end{array}\right) $  in order to form the Lagrangian

\begin{eqnarray}
 L_{e-w} &=&  \frac{i}{2}\hbar c\biggl\{ \overline{\psi}_R \gamma^\mu \p_\mu \psi_R 
    + \overline{\psi}_L \gamma^\mu \p_\mu \psi_L \biggr\} + \mbox{\rm h.c.} \no \\
  & &  \hspace{-.3in}  +  \frac{\hbar c}{4} \biggl\{ \overline{\psi}_R \gamma_5 \hat{\gamma}_\delta \psi_R
         +\overline{\psi}_L \gamma_5 \hat{\gamma}_\delta \psi_L \biggr\}
       \, \epsilon^{\delta\alpha\beta\gamma} e_\alpha^{\,\,\,\lambda} 
             e_\gamma^{\,\,\,\nu} \p_\nu e_{\beta\lambda} 
  + L_{int}
\end{eqnarray}
(plus expressions for the muon and tau). 
$ L_{int} $ contains the electroweak interaction as well as kinetic
terms for $ A_\mu, W^{\pm}_\mu $, and $ Z^0_\mu $.

\newpage

\section*{\large {\bf Appendix C. Proof of \boldmath{$ k^\mu = \alpha K^\mu. $}}}

Begin with equation (49) and substitute the transversality conditions (68) in the form

\begin{eqnarray}
   a^1_{\,\,\,2} & = & \frac{K_3}{2} \biggl\{-\frac{K_1}{K_2 K_3} a^1_{\,\,\,1} - \frac{K_2}{K_3 K_1} a^2_{\,\,\,2} 
          + \frac{K_3}{K_1 K_2} a^3_{\,\,\,3} \biggr\}  \\
   a^2_{\,\,\,3} & = & \frac{K_1}{2} \biggl\{\frac{K_1}{K_2 K_3} a^1_{\,\,\,1} - \frac{K_2}{K_3 K_1} a^2_{\,\,\,2} 
          - \frac{K_3}{K_1 K_2} a^3_{\,\,\,3} \biggr\}  \\
   a^3_{\,\,\,1} & = & \frac{K_2}{2} \biggl\{-\frac{K_1}{K_2 K_3} a^1_{\,\,\,1} + \frac{K_2}{K_3 K_1} a^2_{\,\,\,2} 
          - \frac{K_3}{K_1 K_2} a^3_{\,\,\,3} \biggr\}
\end{eqnarray}
This yields 

\begin{eqnarray}
     \varepsilon^i & = & -\frac{K_0 K^i}{8K_1 K_2  K_3} \biggl\{(K_2^{2} + K_3^{2})
          (a^2_{\,\,\,2} b^3_{\,\,\,3} - a^3_{\,\,\,3} b^2_{\,\,\,2}) \hspace{.75in} \no \\ 
 & & \hspace{-.4in}  + (K_3^{2} + K_1^{2})(a^3_{\,\,\,3} b^1_{\,\,\,1} - a^1_{\,\,\,1} b^3_{\,\,\,3}) 
      +  (K_1^{2} + K_2^{2})(a^1_{\,\,\,1} b^2_{\,\,\,2} - a^2_{\,\,\,2} b^1_{\,\,\,1}) \biggr\} \hspace{.25in} 
\end{eqnarray}
Comparison with  $\var^\mu = (k,\,\, k^ik^0/k) $ shows that $ k^i = \alpha K^i $.  
Next, make use of  $ K_\mu \var^\mu = 0 $ to find 

\begin{eqnarray}
     \varepsilon^0 & = & \frac{(K)^2}{8K_1 K_2  K_3} \biggl\{(K_2^{2} + K_3^{2})
          (a^2_{\,\,\,2} b^3_{\,\,\,3} - a^3_{\,\,\,3} b^2_{\,\,\,2}) \hspace{.55in} \no \\ 
 & & \hspace{-.4in}  + (K_3^{2} + K_1^{2})(a^3_{\,\,\,3} b^1_{\,\,\,1} - a^1_{\,\,\,1} b^3_{\,\,\,3}) 
      +  (K_1^{2} + K_2^{2})(a^1_{\,\,\,1} b^2_{\,\,\,2} - a^2_{\,\,\,2} b^1_{\,\,\,1}) \biggr\} \hspace{.25in} 
\end{eqnarray}
and $ k^0 = \alpha K^0 $.

\newpage

\section*{\large {\bf Appendix D. Gravitational Solution.}}

The transverse, traceless conditions (68) are imposed upon 
the solution (48).
In the equation $ \p_n \xi^n_{\;\; i} = 0, $ the coefficients of $ \cos(-K_\mu x^\mu) $ and
$ \sin(-K_\mu x^\mu) $ must equal zero

\beq
    K_n a^n_{\,\,\,i} = 0  \hspace{.2in} \mbox{\rm and} \hspace{.2in} K_n b^n_{\,\,\,i} = 0
        \hspace{.3in} i = 1,2,3     
\eeq
Similarly, the trace condition $ \xi^n_{\;\; n} = 0 $ requires

\beq 
    a^n_{\;\;n}  = 0  \hspace{.2in} \mbox{\rm and} \hspace{.2in} b^n_{\;\;n}  = 0  
\eeq
These four conditions leave two independent amplitudes 
$ a^i_{\;\;j} $ as well as two $ b^i_{\;\;j} $. [8]

In order to solve the three-dimensional problem, we first choose the independent amplitudes 
to be $ a^1_{\;\; 1},\, a^1_{\;\; 2},\, b^1_{\;\; 1}, $ and $ b^1_{\;\; 2} $.  The  transverse, 
traceless conditions are then used to eliminate all other amplitudes.  Finally, equations 
(77, 78) are used to eliminate $ a^1_{\;\; 2} $ and $ b^1_{\;\; 2} $.  The solution is 

\beq
     \xi^i_{\;\;j} = a^i_{\;\;j} \cos(-K_\mu x^\mu) + b^i_{\;\;j} \sin(-K_\mu x^\mu)
\eeq
with

\begin{eqnarray}
    a^2_{\;\; 2} & = & \frac{1}{(K_2^2 + K_3^2)^2}\biggl\{(K_1^2 K_2^2 - K_3^2 K^2) a^1_{\;\; 1}
            - 2K_1K_2K_3K b^1_{\;\; 1} \biggr\} \hspace{.3in} \\
    a^3_{\;\; 3} & = & \frac{1}{(K_2^2 + K_3^2)^2}\biggl\{(K_1^2 K_3^2 - K_2^2 K^2) a^1_{\;\; 1}
            + 2K_1K_2K_3K b^1_{\;\; 1} \biggr\} \hspace{.3in} \\
    a^1_{\;\; 2} & = & \frac{1}{K_2^2 + K_3^2}\biggl\{- K_1 K_2 a^1_{\;\; 1} + K_3 K b^1_{\;\; 1} \biggr\} \hspace{.3in} \\
    a^2_{\;\; 3} & = & \frac{1}{(K_2^2 + K_3^2)^2}\biggl\{K_2 K_3 (K^2 + K_1^2) a^1_{\;\; 1} 
                      + K_1 K (K_2^2 - K_3^2) b^1_{\;\; 1} \biggr\} \hspace{.3in} \\
    a^3_{\;\; 1} & = & - \frac{1}{K_2^2 + K_3^2}\biggl\{K_1 K_3 a^1_{\;\; 1} + K_2 K b^1_{\;\; 1} \biggr\} \hspace{.3in} 
\end{eqnarray}
and

\begin{eqnarray}
    b^2_{\;\; 2} & = & \frac{1}{(K_2^2 + K_3^2)^2}\biggl\{2K_1K_2K_3K a^1_{\;\; 1} 
                      + (K_1^2 K_2^2 - K_3^2 K^2) b^1_{\;\; 1} \biggr\}\hspace{.3in} \\
    b^3_{\;\; 3} & = & - \frac{1}{(K_2^2 + K_3^2)^2}\biggl\{2K_1K_2K_3K a^1_{\;\; 1}
                      - (K_1^2 K_3^2 - K_2^2 K^2) b^1_{\;\; 1} \biggr\} \hspace{.3in}\\
    b^1_{\;\; 2} & = & -\frac{1}{K_2^2 + K_3^2}\biggl\{K_3 K a^1_{\;\; 1} + K_1 K_2  b^1_{\;\; 1} \biggr\} \hspace{.3in}\\
    b^2_{\;\; 3} & = & -\frac{1}{(K_2^2 + K_3^2)^2}\biggl\{K_1 K(K_2^2 - K_3^2) a^1_{\;\; 1} 
                      - K_2 K_3 (K_1^2 + K^2) b^1_{\;\; 1} \biggr\} \hspace{.3in}\\
    b^3_{\;\; 1} & = & \frac{1}{K_2^2 + K_3^2}\biggl\{K_2 K a^1_{\;\; 1} - K_1 K_3 b^1_{\;\; 1} \biggr\}\hspace{.3in}
\end{eqnarray}

\newpage

\section*{\large {\bf  References.}}

\begin{enumerate}
\item K. Dalton, ``Gravity and the Electroweak Theory,'' 
   \newline www.arxiv.org/physics/0604131.
\item J.J. Sakurai, {\it Advanced Quantum Mechanics}, (Addison-Wesley, 1967) section 3-3.
\item F. Mandl and G. Shaw, {\it Quantum Field Theory}, (Wiley, 1984) app. A.
\item I. Aitchison and A. Hey, {\it Gauge Theories in Particle Physics}, (IOP, 2003) p. 84.
\item B. Thaller, {\it The Dirac Equation}, (Springer, 1992) p. 39.
\item S. Weinberg, {\it Gravitation and Cosmology} (Wiley, 1972) p. 365.
\item V. de Sabbata and M. Gasperini, {\it Introduction to Gravitation} 
  \newline (World Scientific, 1985).
\item C. Misner, K. Thorne, and J. Wheeler, {\it Gravitation}, (Freeman,
    \newline  New York, 1973) section 35.

\end{enumerate}

\end{document}